\documentclass{emulateapj}

\newcommand{\be}{\begin{equation}}
\newcommand{\ee}{\end{equation}}
\newcommand{\ba}{\begin{eqnarray}}
\newcommand{\ea}{\end{eqnarray}}
\newcommand{\bi}{\begin{itemize}}
\newcommand{\ei}{\end{itemize}}
 \newcommand{\bfi}{\begin{figure}
 \epsfxsize=9cm
 \epsffile}
\newcommand{\efi}{\end{figure}}
 \newcommand{\apjl}{\apj}
\newcommand{\mnras}{MNRAS}
\begin{document}

\title{Cosmic dust induced flux fluctuations: bad and good aspects}
\author{Pengjie Zhang$^{1,2}$ and Pier Stefano Corasaniti$^{3,4}$}
\email{pjzhang@shao.ac.cn}
\email{pierste@astro.columbia.edu}
\affil{$^{1}$Shanghai Astronomical Observatory, Chinese Academy of
  Science, 80 Nandan Road, Shanghai, China, 200030}
\affil{$^{2}$Joint Institute for Galaxies and Cosmology (JOINGC) of
SHAO and USTC}

\affil{$^{3}$ISCAP, Columbia University, New York, NY 10027, USA}
\affil{$^{4}$Department of Astronomy, Columbia University, New York, NY 10027, USA}

\begin{abstract}
Cosmic dust extinction alters the flux of type Ia
supernovae. Inhomogeneities in the dust distribution induce
correlated fluctuations of the SN fluxes. We find
that such correlation can be up to $60\%$ of the signal caused by
gravitational lensing magnification, with an opposite sign.
Therefore if not corrected, cosmic dust extinction is
the dominant source of systematic uncertainty
for future SNe Ia lensing measurement
limiting the overall S/N to be $\la 10$.
On the other hand, SN flux correlation measurements
can be used in combination with other lensing data
to infer the level of dust extinction.
This will provide a viable method to eliminate gray dust
contamination from the SN Ia Hubble diagram.
\end{abstract}


\section{Introduction}
Gravitational lensing causes several observable effects
such as distortion of galaxy shape ({\em cosmic shear}),
variation of galaxy number density ({\em cosmic magnification}) and
mode-coupling in cosmic backgrounds. Over the upcoming years measurements of
these effects will provide an accurate mapping of the matter distribution in the universe
(for reviews see \cite{Bartelmann01,Refregier03}).

Recently, several other lensing reconstruction methods have been proposed.
One possibility is to measure the spatial correlation of lensing
induced supernova (SN) flux fluctuations. In fact due to lensing
magnification\footnote{{ Throughout this
paper, the term lensing magnification refers to both the cases of magnification
($\mu>1$) and de-magnification ($\mu<1$). To be more specific, the spatial
correlation functions and the corresponding power spectra ($C_{\kappa}$ and
$C_{\kappa\delta \tau}$) investigated hereafter are averaged over
the full distribution of $\mu$.}}, the SN flux is altered
such that $F\rightarrow F\mu\simeq F(1+2\kappa)$, where $F$ is the
intrinsic SN flux, $\mu$ is the lensing
magnification and $\kappa$ is the lensing convergence.
Intrinsic fluctuations of the
SN flux are random (analogous to intrinsic galaxy ellipticities in cosmic
shear measurement). In contrast those induced by lensing magnification
(see e.g. \citet{Kantowski95,Frieman96,Holz98,Dalal03}) are
correlated with the overall matter distribution (analogous to the
shear signal). Therefore the lensing signature can be inferred either from
spatial correlation measurements of SN fluxes \citep{Cooray06} or from the
root-mean-square of flux fluctuations of high redshift SNe for which the lensing signal is dominant
\citep{Dodelson06}.

Gravitational lensing also induces scatter in the galaxy fundamental
plane through magnification of the effective radius, $R_e\rightarrow
R_e\mu^{1/2}\simeq R_e(1+\kappa)$. Since intrinsic scatters in the
fundamental plane are random, spatial correlation measurements can be used
to infer the lensing signal \citep{Bertin06}. A similar analysis can be applied
to the Tully-Fisher relation as well.

Astrophysical effects may limit the accuracy of these methods. For instance
extinction by cosmic gray dust can be an important source of
systematic uncertainty.
This is because dust absorption changes the apparent SN flux and
may induce correlation of the flux fluctuations. It also
induces scatters in the fundamental plane by dimming the galaxy
surface brightness and
affects the Tully-Fisher relation through dimming the galaxy flux.
These effects potentially cause
non-negligible systematics in the corresponding lensing measurements.

Although the existence of gray dust in the intergalactic medium (IGM) remains
untested, this scenario could account for the metal enrichment of the IGM
(\citet{Bianchi} and reference therein). Testing the gray dust
hypothesis is also relevant
for cosmological parameter inference from SN Ia luminosity distance
measurements.
Recently \citet{Corasaniti06} has pointed out that gray dust models which pass
current astrophysical constraints can induce a $\sim\,20\%$ bias in
the estimate
of the dark energy equation of state $w$
using the Hubble diagram of future SN Ia experiments.

{In this paper, we study the impact of cosmic gray dust on
SN lensing measurements, under the optimistic assumption that
contaminations of reddening dust can be perfectly corrected}. The effects on lensing reconstruction based on
the fundamental plane and the Tully-Fisher relation can be estimated
similarly. For supernova, the key point is that extinction caused by dust
inhomogeneities along the line of sight causes flux fluctuations which
are anti-correlated with the lensing
magnification signal and thus wash-out its imprint.
In particular we find that dust induced correlation can bias SN lensing
measurements by $10-60\%$. Therefore this effect
is likely to be the dominant source of systematics for future SN
surveys characterized by large sky coverage
and sufficiently high surface number density. If not corrected, the
dust induced correlation would limit the signal-to-noise to ${\rm S/N}\la 10$.
This is low compared to the S/N achieved by current cosmic shear measurements
(e.g. \citet{Jarvis05,VanWaerbeke05,Hoekstra05}) and that of proposed
methods such as CMB lensing \citep{Seljak99,Zaldarriaga99,Hu01,Hu02}, 21cm
background lensing \citep{Cooray04,Pen04,Zahn05,Mandel05} and cosmic
magnification of 21cm emitting galaxies \citep{Zhang05,Zhang06}.

Nevertheless we suggest that measurements of the SN flux correlation
still carry valuable information. In fact in combination
with other lensing data they will provide a viable method
to detect and eliminate cosmic gray dust contamination
from future SN Ia luminosity distance measurements.

\section{Dust induced flux Fluctuations}
The observed flux of a SN Ia at redshift $z$ in the
direction $\hat{n}$ of the sky is
\be
F^{\rm obs}(\hat{n},z)=F\mu\,e^{-\tau}\ , \label{flux0}
\ee
where $F$ is the intrinsic flux and $\tau$ the optical depth caused by
dust extinction along the line of sight. The lensing
magnification can be written as $\mu\equiv 1/[(1-\kappa)^2-\gamma^2]\simeq 1+2\kappa$
with $\kappa$ and $\gamma$ being the lensing convergence and
shear respectively. In the presence of dust
density inhomogeneities, the optical depth
can be decomposed in a homogeneous and isotropic part $\bar{\tau}$ and a
fluctuation $\delta{\tau}$ ($\tau\equiv \bar{\tau}+\delta{\tau}$).
To first order, Eq.~(\ref{flux0}) reads as
\be
\label{eqn:flux}
F^{\rm obs}(\hat{n},z)\simeq
Fe^{-\bar{\tau}(z)}\left[1+2\kappa(\hat{n},z)-\delta\tau(\hat{n},z)\right].
\ee
It is worth noticing that the lensing and dust extinction terms have
opposite sign.
Since $\bar{\mu}=0$
(ensemble average)
and $\bar{\kappa}=0$,
the average flux of a SN Ia sample in a given redshift bin is $\bar{F}^{\rm obs}(z)\simeq \bar{F}e^{-\bar{\tau}(z)}$.

The angular correlation of the flux fluctuations can be inferred from
the estimator $\delta_F(\hat{n},z)\equiv F^{\rm obs}/\bar{F}^{\rm
  obs}-1$ \citep{Cooray06}. 
From Eq.~(\ref{eqn:flux}) we then have $\delta_F=2\kappa-\delta\tau$,
hence $\delta_F$ provides an estimate of the gravitational lensing only if
fluctuations in the optical depth are negligible.

The lensing convergence $\kappa$ is related to the 3D matter
overdensity $\delta_m$ by
\be
\label{eqn:kappa}
\kappa=\frac{3}{2}\Omega_m\frac{H_0^2}{c^2}\int \delta_m\,
W(\chi,\chi_s)d\chi\ ,
\ee
where $W(\chi,\chi_s)$ is the lensing
geometry function. For a flat universe $W(\chi,\chi_s)=(1+z)\chi(1-\chi/\chi_s)$,
with $\chi$ and $\chi_s$ the comoving diameter distance to the lens and source
respectively.

Following the derivation of \citet{Corasaniti06} the average optical
depth to redshift $z$ is
\be
\bar{\tau}(z)=\frac{1}{2.5\log{e}}\int_0^z\frac{d\bar{A}}{dz'}c\,dz',
\ee
with $c$ the speed of light and
\be
\frac{1}{2.5\log{e}}\frac{d\bar{A}}{dz}=\frac{3}{4 \varrho}\frac{\bar{\rho}_{\rm d}(z)}{(1+z)H(z)}
\int Q_m^{\lambda}(a,z)N(a)\,\frac{da}{a},
\ee
where $\bar{\rho}_d$ is the average dust density, $\varrho$ is the
grain material density, $a$ is the grain size,
$Q_m^{\lambda}$ is the extinction efficiency factor at the rest-frame
wavelength $\lambda$ which depends on the grain size and complex refractive
index $m$, and $N(a)$ is the size distribution of dust particles.
{The extinction efficiency factor is computed by solving numerically the
Mie equations for spherical grains (Barber \& Hill 1990).
Since dust particles are made of metals we estimate the evolution of
the average cosmic dust density $\bar{\rho}_d$ from the
redshift dependence of the average cosmic metallicity as
inferred by integrating the star formation history (SFH) the Universe.
Such a modeling is an extention of that presented in Aguirre (1999)
and Aguirre \& Haiman (2000), since in addition to estimating the amount of
cosmic dust density in terms of the measured SFH, it accounts for the physical
and optical properties of the dust grains.

This approach differs from that used in some of the SN Ia literature
(see for instance Riess et al. (2004)). In these studies the cosmic
dust dimming is estimated by modeling the evolution of dust density
as a redshift power law with different slopes corresponding
to different cosmic dust models. More importantly these studies assume
the empirical interstellar extinction law, tipically in the form inferred
by Cardelli et al. (1989). However cosmic dust particles undergo
very different selection mechanisms compared to interstellar grains
and therefore are unlikely to cause a similar extinction.

\bfi{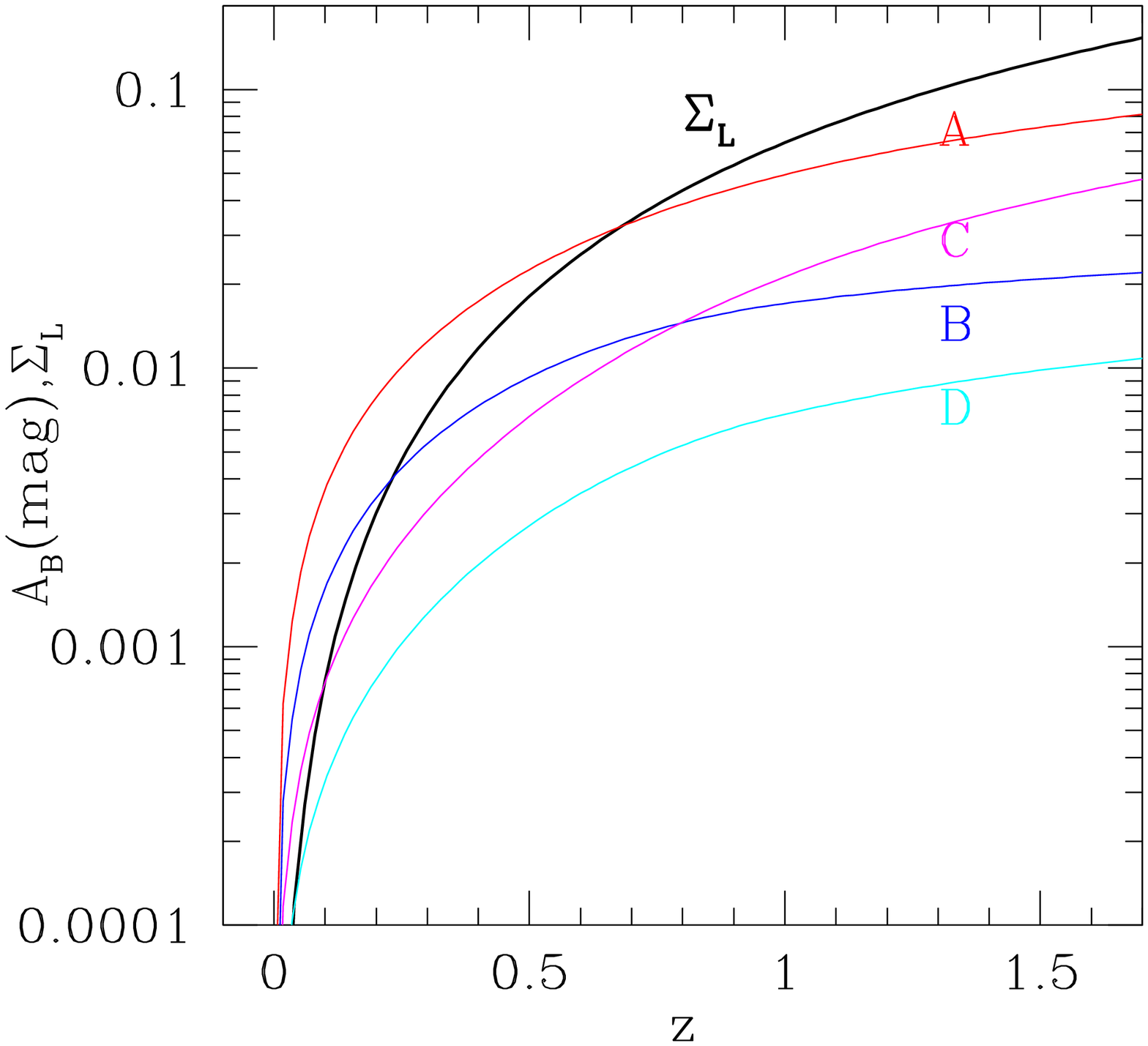}
\caption{The lensing normalized matter surface density $\Sigma_L$
and the B-band dust extinction $A_B$ for different dust models (see text).
Since $A_B$ and $\Sigma_L$ are comparable, dust extinction effects cannot be
neglected in lensing measurements of SN flux correlation.   \label{fig:A}}
\efi

In this perspective our modeling is rather robust, since the cosmic dust
absorption is computed from first principles and in terms of astrophysical
parameters which can be measured through several observations, such
as X-ray quasar halo scattering (see Paerels et al. 2002) or high resolution
measurements of the Far Infrared Background (FIRB) (Aguirre \& Haiman 2000). For more details on these cosmic dust models and their cosmological
impact we refer to \citet{Corasaniti06}.}

The fluctuation in the optical depth is then given by
\be
\delta{\tau}=\frac{1}{2.5\log{e}}\int_0^z \frac{d\bar{A}}{dz'}\delta_{\rm d}(z') c\,dz',
\ee
where $\delta_{\rm d}$ is the fractional dust density perturbation.
The resulting auto-correlation power spectrum of $\delta_F$ is:
\be
\frac{1}{4}C_{\delta_F}(l)=C_{\kappa}+\frac{1}{4}C_{\delta\tau}-C_{\kappa\delta{\tau}},\label{cl}
\ee
where $C_{\kappa}$, $C_{\delta\tau}$, $C_{\kappa\delta{\tau}}$ are
the angular power spectra of $\kappa$, $\delta{\tau}$, and the
$\kappa$-$\delta{\tau}$ cross correlation.
Using the Limber's approximation these read as \citep{Limber54,Kaiser98}:
\be
\frac{l^2C_{\kappa}}{2\pi}=\frac{\pi}{l}
  \left[\frac{3\Omega_mH_0^2}{2c^2}\right]^2 \int
  \Delta^2_{\delta}\left(\frac{l}\chi,z\right) W^2(\chi,\chi_s)\chi d\chi\ ,
\ee
\be
\frac{l^2C_{\delta {\tau}}}{2\pi}=\frac{\pi}{l}
  \left[\frac{1}{2.5\log e}\right]^2 \int
  \Delta^2_{\delta_d}\left(\frac{l}\chi,z\right)
  \left[\frac{d\bar{A}}{d\chi}\right]^2\chi d\chi \ ,
\ee
and
\be
\frac{l^2C_{\kappa\delta{\tau}}}{2\pi}=\frac{\pi}{l}
  \frac{3\Omega_mH_0^2}{5c^2\log e}\int
  \Delta^2_{\delta\delta_d}\left(\frac{l}\chi,z\right) W(\chi,\chi_s)\frac{d\bar{A}}{d\chi}\chi d\chi,
\ee
where $\Delta^2_{\delta}\equiv k^3P_{\delta}(k)/2\pi^2$ is the
dimensionless matter
density variance and $P_{\delta}$ is the matter density power
spectrum. The nonlinear
$\Delta^2_{\delta}$ is
calculated using the Peacock-Dodds fitting
formula \citep{Peacock96}. $\Delta^2_{\delta\delta_d}$ and
$\Delta^2_{\delta_d}$ are defined analogously.
The spatial distribution of IGM dust is not known, the simplest assumption
is that dust traces the total mass distribution. In such
case $\Delta^2_{\delta_d}=b_d^2\Delta^2_{\delta}$ and
$\Delta^2_{\delta\delta_d}=b_d\Delta^2_{\delta}$, where
$b_d$ is the dust bias.

Defining $\Sigma_L\equiv \frac{3}{2}\Omega_m\frac{H_0^2}{c^2}\int W(\chi,\chi_s)d\chi$,
one has $\delta{\tau}/\kappa\sim b_d\bar{A}/\Sigma_L$, hence
$C_{\delta{\tau}}/C_{\kappa}\sim b_d^2(\bar{A}/\Sigma_L)^2$
and $C_{\kappa\delta{\tau}}/C_{\kappa}\sim b_d(\bar{A}/\Sigma_L)$.
This indicates that cosmic dust contamination is negligible only if
$\bar{A}(z)\ll \Sigma_L(z)$.

We adopt a flat $\Lambda$CDM cosmology with $\Omega_m=0.3$,
$\Omega_{\Lambda}=0.7$, $h=0.7$, $\Omega_b=0.04$, $\sigma_8=0.9$ and
the primordial power index $n=1$.  We assume the BBKS transfer function
\citep{BBKS}. {For the dust extinction we limit our analysis to a test-bed of four cosmic dust models
studied in \citet{Corasaniti06}. These are characterized by model parameter values motivated by
astrophysical considerations. In particular the particle size distribution is
in the range $0.05-0.2\,\mu{m}$, consistently with the fact that smaller grains are distroyed by
sputtering, while larger ones remain trapped in the gravitational
potential of the host galaxy
(Ferrara et al. (1991); Shustov \& Vibe (1995); Davis et al. (1998)).
The grain composition consisting of Silicate or Graphite particles and we
consider both low and high star
formation history scenarios.

Models A and B refer to Graphite particles with low and high SFH respectively,
while models C and D correspond to Silicate grains. The total dust density for these models
is within the
limits imposed by the DIRBE/FIRAS data (Aguirre \& Haiman 2000) and concides
with the upper limit obtained
from the analysis of X-ray quasar halo scattering (Paerels et
al. 2002). These gray dust models
cause little reddening of the incoming light and induce a color excess
in the optical
and near-IR bands smaller than $0.01\,{\rm mag}$}.

\bfi{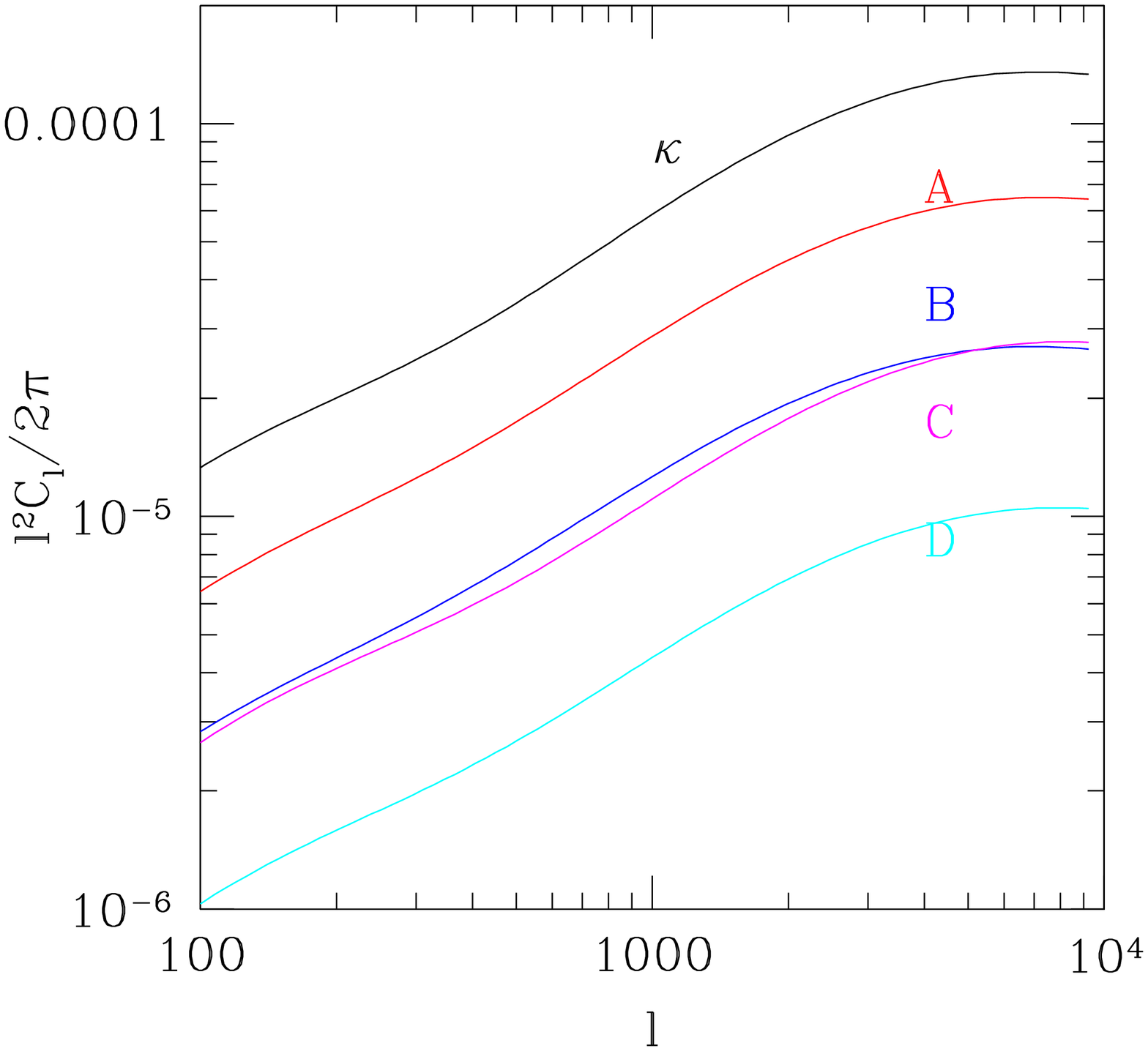}
\caption{Lensing and dust contamination power spectra. The upper line is
the lensing convergence $l^2C_{\kappa}/2\pi$. Other lines are
$C_{\kappa\delta{\tau}}-C_{\delta{\tau}}/4$ for dust model
A, B, C and D, respectively with $b_d=1$. We have assumed all SNe to be at
$z_s=1$. Clearly, the existence of cosmic dust would degrade
or even prohibit measurement of the lensing signal. \label{fig:cl}
}
\efi

{A further assumption
concerns the gray dust spatial distribution, for which we  have little
knowledge of. This model uncertainty may affect the results presented
in this paper significantly. One can imagine an extreme case where
gray dust distributes homogeneously. Then there will be no fluctuations in
$\tau$ and thus no induced correlation in SN flux
fluctuations. However, current understanding of gray dust
formation implies that gray dust is associated with
the overall matter distribution. So a more
appropriate treatment of gray dust distribution is the bias model
$\delta_d=b_d\delta_m$, as adopted in this paper. Although it is
natural to expect $b_d$ to be redshift
and scale dependent, since we have little knowledge of it for
simplicity we assume $b_d=1$.}

From Fig.~\ref{fig:A} we can see that $\Sigma_L$
is comparable to the B-band dust extinction $\bar{A}$, hence dust
contamination cannot be neglected.
Therefore $C_{\delta{\tau}}$ and $C_{\kappa\delta{\tau}}$ in
Eq.~(\ref{cl}) are source of systematic errors which need
to be corrected if we aim to measure the convergence power
spectrum.

In Fig.~\ref{fig:cl} we plot the lensing convergence power spectrum
$l^2C_\kappa/2\pi$, and the dust contamination
power spectrum $C_{\kappa\delta\tau}-C_{\delta\tau}/4$ for our test-bed of dust models
for sources at $z_s=1$. We find that $C_{\delta{\tau}}$ is smaller
than $C_{\kappa\delta{\tau}}$, mainly due to the $1/4$ prefactor.
Since  $C_{\kappa\delta{\tau}}$ has an opposite sign to the lensing
signal in Eq.~(\ref{cl}), its overall effect is to suppress the
spatial correlation of SN Ia flux fluctuations and consequently
diminish the variance and covariance of flux fluctuations.  Since
statistical errors on cosmological parameters constraints from SNe
Ia Hubble diagram are proportional to the square root of the variance
and covariance (see e.g. \citet{Cooray06a} for discussions), the
existence of cosmic dust extinction fluctuations decreases the
statistical uncertainties, though the mean dust extinction
will induce a systematic bias unless corrected.

Dust contamination can be quantified by the ratio
$\eta\equiv |C_{\kappa\delta{\tau}}-C_{\delta{\tau}}/4|/C_{\kappa}$. Since
both $\kappa$ and $\delta \tau$ trace the same large scale structure
(enforced by the simplification $b_d=$const.),
the multipole
dependence of $C_\kappa$, $C_{\delta\tau}$ and
$C_{\kappa\delta\tau}$ are similar such that $\eta$ is
roughly constant. In table~\ref{table:error} we list its values for sources at redshift $z_s=0.5,1.0$ and $1.7$
respectively. As it can be seen, model
A causes the largest contamination inducing
a systematic error as large as $60\%$ of lensing signal. Even for model D the contamination is still $\sim 10\%$,
which is comparable to the statistical error expected from future SN Ia
lensing measurements. Consequently dust induced systematics
will be the dominant source of uncertainty for this type of measurements.

Furthermore we find that the relative error can be approximated by $\eta=\beta b_d\bar{A}/\Sigma_L$,
where $\beta\simeq 0.7$ with a dispersion $<0.1$ over the redshifts investigated for our test-bed
of dust models. This relation suggests that if we can measure $\eta$ in combination with an
independent lensing measurement, it would be possible to infer $ \bar{A}$ given knowledge of $b_d$.
In the next section we will discuss how these type of measurements can be used to
remove cosmic dust contamination in the SN Ia Hubble diagram.

\section{Removing cosmic dust contamination}
\label{sec:discussion}
{Flux fluctuations induced by lensing and extinction are
small compared to intrinsic SN flux fluctuations and therefore can only
be extracted statistically, except for the strongly lensed or heavily
extincted SNe.}
Accurate lensing measurements can be obtained from a variety of
astrophysical observations of cosmic shear and cosmic
magnification. In combination with correlation measurements of SN fluxes these
can be used to quantify the level of cosmic dust extinction and
provide a viable method to remove dust systematics from the SN Ia Hubble diagram.
The idea is to infer $\eta$ from the comparison
of $C_{\delta_F}$ and $C_{\kappa}$.
As discussed before $\eta\simeq 0.7 b_d\bar{A}/\Sigma_L$,
this would allow to measure  $\bar{A}$ up to model uncertainties in $b_d$
and measurement errors in $C_{\delta_F}$. The estimated value
of $\bar{A}$ can then be used to correct the standard-candle
relation of SN Ia.

\begin{table}
\begin{center}
\caption{The relative error caused by extinction with respect to
  lensing, $\eta=|C_{\kappa\delta{\tau}}-C_{\delta{\tau}}/4|/C_{\kappa}$
at different source redshifts for our test-bed of dust models.
$\eta$ is roughly independent of  multipole $l$, since shapes of $C_{\kappa}$,
  $C_{\kappa\delta{\tau}}$ and $C_{\delta{\tau}}$ are very
  similar. \label{table:error}}
\begin{tabular}{ccccc}
\hline
\hline
source redshift &Graphite& Graphite& Silicate& Silicate\\
$z_s$&high SFH&low SFH& high SFH& low SFH \\
&Model A & Model B & Model C & Model D\\
\hline
0.5&0.65 &0.33 &0.24 &0.11\\
1.0&0.45&0.20&0.21&0.08\\
1.7&0.36 &0.13 & 0.19& 0.06\\
\hline
\hline
\end{tabular}
\end{center}
\end{table}

The efficiency of this method depends on the sky coverage and the SN number
density of the survey. For instance in order to measure
$\bar{A}$ to $10\%$ accuracy, the overall S/N of $C_{\delta_F}$
must be $\ga 10(1/\eta-1)$. This implies that for model A,
$C_{\delta_F}$ should be measured with S/N of $\sim 10$, while for model B, C and D,
it would require a $S/N\geq 40$-$100$. In the case of a survey with
$10^4$ SNe and covering $20$ deg$^2$ of the sky the signal-to-noise is $S/N\sim 10$ \citep{Cooray06}.
Since $S/N\propto f_{sky}^{1/2}$ to reach $S/N=40$-$100$ requires
a factor of 20-100 times higher in sky coverage and total number of observed SNe.
This can be achieved by the proposed ALPACA experiment \citep{ALPACA}.

Galaxy-quasar correlation measurements provide another method to
estimate the level of cosmic dust extinction. For a given line of
sight, dust extinction reduces the observed number of galaxies
above flux $F$ from $N(>F)$ to $N(>F\exp[\bar{\tau}+\delta
  \tau])\simeq N(>F)[1-\alpha (\bar{\tau}+\delta{\tau})]$.  Here, $\alpha=-d\ln
N/d\ln F$ is the (negative) slope of the intrinsic galaxy luminosity function $N(>F)$
and we have assumed $\tau\ll 1$. Thus dust inhomogeneities induce
a fractional fluctuation $-\alpha\delta{\tau}$ in the galaxy number density.
Since $\delta{\tau}$ is correlated with the matter density field,
dust extinction induces a correlation between foreground
galaxies and background galaxies (quasars) such that
$w_{fb}(\theta)=-\alpha\langle \delta{\tau}({\bf
  \theta^{'}})\delta_g^f({\bf\theta}^{'}+{\bf \theta})\rangle$. Here,
$\delta^f_g$ is the foreground galaxy number overdensity.
On the other hand, lensing induced fluctuations in galaxy number
density are $2(\alpha-1)\kappa$ \citep{Bartelmann01}, where the $-1$ term
accounts for the fact that lensing magnifies the surface area and thus decreases
the number density. Because of the different dependence on the slope $\alpha$
the signal of extinction and lensing can be separated simultaneously.

The SDSS galaxy-quasar cross correlation measurement
\citep{Scranton05} is consistent with the $\alpha-1$ scaling and
thus the dust contamination, if any, remains sub-dominant. Our dust
models are consistent with this
measurement, since the
expected fractional contribution from dust extinction is
$-\alpha/[2(\alpha-1)]\langle\delta{\tau}\delta^f_g\rangle/\langle\kappa\delta^f_g\rangle\sim
-\alpha/(\alpha-1)\times(0.27,0.11,0.08,0.03)$ for dust model A, B, C
and D respectively. However, such measurement is already at the edge
of providing interesting dust constraints. For instance, model A
induces at $\theta=0.01^{\circ}$ a negative correlation with amplitude $\sim
0.003\alpha b_g$, where $b_g$ is the SDSS galaxy bias.  This signal is
already very close  to the
measurement uncertainty (Fig.7, \citet{Scranton05} and the averaged
$\langle \alpha\rangle\simeq 1$ from their table 2). {In principle,
by combining color and flux dependences of the galaxy-quasar
cross correlation and the color-galaxy  cross correlation, it will
be possible to separate the contribution of lensing
magnification, gray and reddening simultaneously
\footnote{Private communication with Brice Menard. }.}
The next generation of galaxy surveys such as LSST, ALPACA or PanSTARRS
will provide foreground galaxy-quasars measurements that can achieve
a S/N $\gg 10$. This will allow to discriminate the above dust
models unambiguously, thus providing accurate constraints on the cosmic
dust extinction and clustering properties.

\section{Conclusions}
Several new methods have been proposed for inferring the lensing magnification
signal from a variety of correlation measurements. These involve SN Ia flux,
the fundamental plane and Tully-Fisher relation of optical galaxies.
In this paper we have shown that contamination of cosmic dust
extinction may severely degrade such measurements. As an example
inhomogeneities in the cosmic dust distribution
may limit the S/N of SN lensing measurements to $\la 10$ level.

Billions of galaxies can be  detected/resolved by  the Square Kilometer
Array\footnote{SKA: http://www.skatelescope.org/} through the 21cm
hyperfine transition line emission which is not affected by
dust extinction. In such a case the only scatters other than intrinsic
ones in the Tully-Fisher relation ($L\propto v_c^4$) are induced by lensing magnification,
$L\rightarrow L(1+2\kappa)$. Therefore lensing reconstruction using these
galaxies is an attractive possibility, since it is free of some systematics
associated with cosmic shear such as shape distortion induced by point
spread function.

On the other hand, spatial correlation measurements of SN fluxes or
galaxy-quasar will
constrain the amount of cosmic gray dust and its
clustering properties to high accuracy. This will not only provide a
better understanding of IGM dust physics, but also
a valuable handing of dust contamination in the SN Ia Hubble diagram.

{\it Acknowledgments}---. We thank Brice Menard and Ryan Scranton for
helpful discussions on dust contamination in SDSS samples. We
are also thankful to Yipeng Jing, Peter Nugent and Alexandre Refregier
for useful discussions.   
PJZ is supported  by the One-Hundred-Talents Program of Chinese Academy
of Science   and the NSFC grants
(No. 10543004, 10533030).


\begin{thebibliography}{}
\bibitem[Aguirre (1999)]{Aguirre99} Aguirre, A., 1999, \apj, 525, 583
\bibitem[Aguirre \& Haiman (2000)]{AgHa} Aguirre, A., Haiman, Z., 2000, \apj, 532, 28
\bibitem[Barber \& Hill (1990)]{Barber} Barber, P.~W., Hill, S.~C., 1990, Light Scattering
by Particles: Computational Methods, World Scientific Publishing, Singapore
\bibitem[Bardeen et al.(1986)]{BBKS} Bardeen, J.~M., Bond, J.~R., Kaiser, N., Szalay, A.~S. 1986, \apj, 304, 15
\bibitem[Bartelmann \& Schneider(2001)]{Bartelmann01} Bartelmann, M.,  Schneider, P. 2001, Phys. Rept., 340, 291
\bibitem[Bertin \& Lombardi(2006)]{Bertin06} Bertin, G., Lombardi, M.\ 2006, astro-ph/0606672
\bibitem[Bianchi \& Ferrara(2005)]{Bianchi} Bianchi, S., Ferrara, A. 2006, Mont. Not. Roy. Astron. Soc., 358, 379. \apj, 345, 245
\bibitem[Cardelli et al. (1989)]{Card89} Cardelli, J.~A., Clayton, G.~C., Mathis, J.~S., 1989,
\bibitem[Cooray(2004)]{Cooray04} Cooray, A. 2004, New Astron. Rev. 9, 173
\bibitem[Cooray et al.(2006a)]{Cooray06a} Cooray, A., Huterer, D., Holz, D.~E. 2006, Phys. Rev. Lett., 96, 021301
\bibitem[Cooray et al.(2006)]{Cooray06} Cooray, A., Holz, D.~E., Huterer, D. 2006, \apj, 637, L77
\bibitem[Corasaniti(2006)]{Corasaniti06} Corasaniti, P. S., 2006, Mont. Not.
Roy. Astron. Soc., 372, 191
\bibitem[Corasaniti et al.(2006)]{ALPACA} Corasaniti, P. S., LoVerde, M., Blake, C., Crotts, A. 2006,
Mont. Not. Roy. Astron. Soc., 369, 798
\bibitem[Dalal et al.(2003)]{Dalal03} Dalal, N., Holz, D.~E., Chen, X., Frieman, J.~A. 2003, \apjl, 585, L11
\bibitem[Davies et al. (1998)]{Dav98} Davies, J.~I., Alton, P.~B., Bianchi, S., Trewhella, M., 1998,
\mnras, 300, 1006
\bibitem[Dodelson \& Vallinotto(2006)]{Dodelson06} Dodelson, S., Vallinotto, A. 2006, astro-ph/0511086
\bibitem[Ferrara et al. (1991)]{Ferr91} Ferrara, A., Ferrini, F., Barsella, B., Franco, J., 1991,
\apj, 381, 137
\bibitem[Frieman (1996)]{Frieman96} Frieman,J. 1996, astro-ph/9608068
\bibitem[Hoekstra et al.(2005)]{Hoekstra05} Hoekstra, H., et al. 2005, astro-ph/0511089
\bibitem[Holz (1998)]{Holz98} Holz, D.~E. 1998, \apjl, 506, L1
\bibitem[Hu (2001)]{Hu01} Hu, W. 2001, \prd, 64, 083005
\bibitem[Hu \& Okamoto(2002)]{Hu02} Hu, W., Okamoto, T. 2002, \apj, 574, 566
\bibitem[Jarvis et al. (2005)]{Jarvis05} Jarvis, M., Jain, B., Bernstein, G., Dolney, D. 2005, astro-ph/0502243
\bibitem[Kaiser (1998)]{Kaiser98} Kaiser, N. 1998, \apj, 498, 26
\bibitem[Kantowski et al. (1995)]{Kantowski95} Kantowski, R., Vaughan, T., Branch, D. 1995, \apj, 447, 35
\bibitem[Limber (1954)]{Limber54} Limber, D.~N. 1954, \apj, 119, 655
\bibitem[Mandel \& Zaldarriaga(2005)]{Mandel05} Mandel, K.S., Zaldarriaga, M. 2005, astro-ph/0512218
\bibitem[Paerels et al. (2002)]{Paerels} Paerels, F., Petric, A., Telis, A., Helfand, D.~J., 2002,
BAAS, 201, 97.03
\bibitem[Peacock \& Dodds(1996)]{Peacock96} Peacock, J.~A., Dodds, S.~J. 1996, Mont. Not. Roy. Astron. Soc., 280, L19
\bibitem[Pen(2004)]{Pen04} Pen, U. 2004, New Astronomy, 9, 417
\bibitem[Refregier(2003)]{Refregier03} Refregier, A. 2003, ARAA, 41, 645
\bibitem[Riess et al. (2004)]{Riess04} Riess, A.~G. et al., 2004, \apj, 607, 665
\bibitem[Scranton et al. (2005)]{Scranton05} Scranton, R., et al. 2005, \apj, 633, 589
\bibitem[Seljak \& Zaldarriaga (1999)]{Seljak99} Seljak, U., M. Zaldarriaga 1999, Phys. Rev. Lett., 82, 2636
\bibitem[Shustov \& Vibe (1995)]{Shu95} Shustov, B.~M., Vibe, D.~Z., 1995, Astronomy Reports, 39, 579
\bibitem[Van Waerbeke et al. (2005)]{VanWaerbeke05} Van Waerbeke, L., Mellier, Y., Hoekstra, H. 2005, \aap, 429, 75
\bibitem[Zahn \& Zaldarriaga (2005)]{Zahn05} Zahn, O., Zaldarriaga, M., 2005, astro-ph/0511547
\bibitem[Zaldarriaga \& Seljak (1999)]{Zaldarriaga99} Zaldarriaga, M., Seljak, U. 1999, Phys. Rev. D, 59, 13507
\bibitem[Zhang \& Pen (2005)]{Zhang05}  Zhang, P.J., Pen, U.L. 2005, \prl,  95, 241302
\bibitem[Zhang \& Pen (2006)]{Zhang06} Zhang, P.J., Pen, U.L. 2006, Mont. Not. Roy. Astron. Soc., 367, 169
\end{thebibliography}
\end{document}